\documentclass{aa}

\usepackage{graphicx}

\begin{document}



\title{Properties of RR Lyrae stars in the Inner Regions of the Large Magellanic Cloud.
\thanks{Based on observations collected with the 
Very Large Telescope and the New Technology Telescope
of the European Southern Observatory within the Observing Programs
64.N-0176(B) and 70.B-0547. Tables 3, 4 and 6 are also available in electronic form
at the CDS via anonymous ftp to cdsarc.u-strasbg.fr (130.79.128.5)
or via http://cdsweb.u-strasbg.fr/cgi-bin/qcat?J/A+A/}}

\author{J.~ Borissova\inst{1}
\and
D.~Minniti\inst{1}
\and
M.~ Rejkuba\inst{2}
\and 
D.~Alves\inst{3}
\and
 K. H. ~Cook\inst{4}
\and 
K. C. ~Freeman\inst{5}. 
}

\offprints{J.~Borissova}

\institute{
      Department of Astronomy, P. Universidad Cat\'olica, 
      Av. Vicu\~na Mackenna 4860, Casilla 306, Santiago 22, Chile\\
      \email{jborisso, dante@astro.puc.cl}
\and
      European Southern Observatory, Karl-Schwarzschild-Str. 2, D-85748
 Garching b.  M\"{u}nchen, Germany\\
      \email{mrejkuba@eso.org}
\and
      Columbia University, Dept. of Astronomy, New York, USA\\
      \email{alves@astro.columbia.edu}
\and
      Lawrence Livermore National Laboratory, Livermore, California, USA\\
      \email{kcook@llnl.org}
\and
      Mount Stromlo Observatory, Canberra ACT, Australia\\
      \email{kcf@mso.anu.edu}
}

\date{Received .. ... 2003; accepted .. ... 2003}

\authorrunning{Borissova et al.}
\titlerunning{ Properties of LMC RR Lyrae stars}

\abstract{ 
We present the radial velocities, metallicities and 
the K-band magnitudes of 74 RR Lyrae stars in the inner regions of
the LMC. The intermediate resolution spectra and infrared images were obtained 
with FORS1 at the ESO VLT and with the SOFI infrared imager at the ESO NTT.  
The best 43 RR Lyrae with measured velocities yield an observed velocity
dispersion of $\sigma=61\pm 7$ km s$^{-1}$. We obtain a true LMC RR Lyrae 
velocity dispersion of $\sigma=53$ km s$^{-1}$, which is higher than the 
velocity dispersion of any other LMC population previously measured.  
This is the first empirical evidence for a  kinematically hot, 
metal-poor halo in the LMC as discussed in Minniti et al. (2003). 
Using Layden's (1994) modification of the $\Delta S$ method 
we measured the metallicity for 23 of our stars. The mean value is
[Fe/H] =$-1.46\pm0.09$ dex. The absolute magnitudes $M_{V}$ 
and $M_{K}$ of RR Lyrae stars are linear functions 
of metallicity. In the V band, our data agree with the 
Olech et al. (2003) relation, in the K band the slope is flatter.  
The average apparent V luminosity of 70 RR Lyrae stars is $<V> = 19.45\pm0.04$ 
and the average K luminosity of 37 RR Lyrae stars is $<K>=18.20\pm0.06$.  
There is no obvious relation between apparent V magnitude and LogP, 
while the RR Lyrae K band magnitudes show a well defined linear trend
with LogP.
Using the Bono et al. (2001) and Bono et al. (2003) theoretical 
Near-Infrared Period-Luminosity-Metallicity relations  
we calculate the LMC distance modulus $\mu_{0} = 18.48\pm0.08$.

%
%

}

\maketitle

\section{Introduction}

RR~Lyrae stars are old 
($t \ga 10^{10}$ yr), 
metal-poor 
($-2.5< \mathrm{[Fe/H]}<-0.5$)
pulsating variables.
They are one of the most important steps in the distance ladder, and are
also one of the best tracers of the primordial populations of the Milky Way.
Their characteristic light curves make them easy to find in large microlensing
surveys like the MACHO Project.

Studies of the LMC kinematics (Alves \& Nelson 2000, Graff et al. 2000,
Gyuk et al. 2000, Hardy et al. 2001, van der Marel et al. 2002)
suggest that if they lie in a disk, like
the globular clusters, the RR Lyrae stars would show fairly rapid
systemic rotation (about $\pm40$ km s$^{-1}$), and a velocity
dispersion of about $15-30$ km s$^{-1}$.
If they belong to a metal-poor halo, then the rotation will be low and the
velocity dispersion will be about $50$ km s$^{-1}$.

Walker (1992) measured the abundances of 182 RR Lyrae stars from 
the period-amplitude
relation in seven LMC clusters  (NGC\,2257, Reticulum, NGC\,1841, NGC\,1466,
NGC\,1766, NGC\,2210, NGC\,1835). They fall in the interval -1.7 to -2.3, with
the mean value [Fe/H]=$-1.9\pm0.2$ dex.  
Recently, Clementini et al. (2003)  measured metallicities of 101 
LMC field, RR Lyrae stars using low-resolution spectra 
obtained with the Very Large Telescope. They found metallicities 
between -0.5 and -2.1, with
an average value of [Fe/H]=$-1.46\pm0.3$ dex. 

The distance to the LMC has been the subject of many studies. 
There is a well known discrepancy of 0.2---0.3 mag between distances
derived from Population I and II indicators. 
Population I distance indicators give a long distance modulus for the LMC, 
in the range from 18.5 to 18.7 mag, while
Population II indicators  support a shorter
modulus in the range from 18.4 to 18.6.
Clementini et al. (2003) summarized most of the distance determinations and
using accurate photometry of 101 RR Lyrae stars, reported a common value of
$\mu=18.515\pm 0.085$ mag.

The goal of this project is to measure the properties (kinematics,
distributions, abundances and distances) of the RR Lyrae stars 
in six selected inner region fields of the LMC.
To use the LMC RR Lyrae as tracers of a putative LMC halo is not a new idea.
Kinman et al. (1991), and Feast (1992) review the problem of measuring their
kinematics. Because these LMC RR Lyrae stars are faint ($19<V<20$),
ten years ago this project was beyond the reach of the best telescopes.
Metallicities and distance determination are additional products of 
this project.

\section{Observations and Reductions}

\subsection{Selection of the sample}
We have chosen our sample of
RR Lyrae stars using the MACHO database
from six central fields of the LMC bar,
at distances from 0.7 to 1.5 degrees away from the rotation center.
As a control sample for 
the 
LMC kinematic properties, we also selected
known long period variables (LPV)  from MACHO and OGLE (Zebrun et al. 2001)
and Cepheids
from the MACHO catalog
in the same fields. The observed fields are shown on Fig.~\ref{Fig01}
and are summarized in Table~1. The fields are centered on the source stars
of the LMC microlensing events discovered by MACHO Project (Alcock et al. 2001).
The number of stars for each field with spectra and/or K band magnitudes
is also given in the last column.

\begin{table*}[t]\tabcolsep=0.1pt\tiny
\begin{center}
\caption{The centers of the observed fields. The Log of the observations.}
\label{Table1}
\begin{tabular}{l@{\hspace{5pt}}l@{\hspace{5pt}}l@{\hspace{7pt}}rcccccc}
\hline
\multicolumn{1}{l}{Event}{\hspace{5pt}}&
 \multicolumn{1}{c}{RA}&
\multicolumn{1}{c}{DEC}{\hspace{3pt}}&
\multicolumn{1}{c}{MJD(SOFI)}{\hspace{3pt}}&
\multicolumn{1}{c}{ExpTime (min)}{\hspace{3pt}}&
\multicolumn{1}{c}{No of Exp.}{\hspace{3pt}}&
\multicolumn{1}{c}{Date(FOR1)}{\hspace{3pt}}&
\multicolumn{1}{c}{ExpTime (min)}{\hspace{3pt}}&
\multicolumn{1}{c}{No of Exp.}{\hspace{3pt}}&
\multicolumn{1}{c}{RRLyr's}\\
\hline
LMC-1   &   05:14:44.3 & -68:48:01&\hspace{3pt}51594.107748&20&2&10.1.2003&20&2& 23\\
LMC-4   &   05:17:14.6 & -70:46:59&\hspace{3pt}51594.129148&20&2&11.1.2003&20&2& 13\\
LMC-7   &   05:04:03.4 & -69:33:19&\hspace{3pt}51594.204557&20&2&11.1.2003&20&2& 5 \\
LMC-9   &   05:20:20.3 & -69:15:12&\hspace{3pt}51594.172106&20&2&11.1.2003&20&2& 13\\
LMC-12  &   05:33:51.7 & -70:50:59&\hspace{3pt}51594.214012&20&2&12.1.2003&20&2& 8\\
LMC-14  &   05:34:44.0 & -70:25:07&\hspace{3pt}51594.254683&20&2&12.1.2003&20&2&13\\   
\hline
\end{tabular}
\end{center}
\end{table*}

%
\begin{figure}[h]
\caption{ This is MACHO image (http://www.macho.mcmaster. ca/)
  of LMC (R-band) showing
	the location of the MACHO fields and the fields for which 
	we obtained MOS spectra and K band photometry. 
	North is up and east is to the left. The 
	small thick 
	boxes indicate the  $7\arcmin \times\ 7\arcmin$ field of view of
	the CCD images.
	}
 \label{Fig01}
\end{figure}

\begin{figure}[h]
\resizebox{\hsize}{!}{\includegraphics{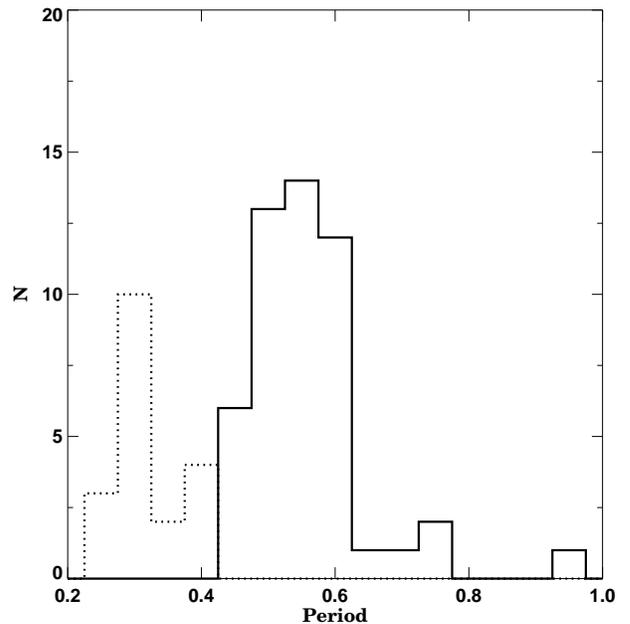}}
 \caption{ The period histogram for RR Lyrae stars in our fields.
The solid line stand for RRab stars, RRc and RRe stars are marked 
with a dotted line.
}
 \label{Fig02}
\end{figure}

The distribution over the period of our RR Lyrae stars is shown 
on Fig.~\ref{Fig02}. The solid line represents RRab stars, while the
dotted line marks RRc+RRe stars.
Our sample of RRab stars covers the interval from
0.35 days to the 0.75 days (only one star has period of 0.97 days), 
with a mean value of $0.572\pm0.09$ days, 
the RRc stars range from 0.25 to 0.43 days
and the mean period is $0.352\pm0.05$ days.
The mean periods of our RRab Lyrae sample is in good agreement with values
determined by the MACHO team from approx. 7900 RR Lyrae stars:
${\langle P_{ab}\rangle=0.583}$ days (Alcock et al. 1996)
and the recent work of the
OGLE team (Soszynski et al. 2003): ${\langle P_{ab}\rangle=0.573}$
days. For RRc stars the mean 
periods are ${\langle P_c\rangle=0.342}$ days, from the MACHO team, and 
${\langle P_c\rangle=0.339}$, from the OGLE team.
Therefore, despite the limited sample number, the properties 
of RR Lyrae stars in
our sample should be representative of the whole population.

Soszynski et al. (2003) present the period distribution for OGLE RR Lyrae
in the LMC fields and in old LMC globular clusters. Their period 
distribution and mean periods of field LMC RR~Lyrae 
agree with those measured by
Alcock et al. (2000).  However, we note that
the LMC globular cluster RR Lyrae show a different period distribution,
peaked at shorter periods both for RRab and RRc stars. 
This difference is worth noticing, because
it suggests that globular clusters and field RR Lyrae in the LMC do
not trace similar populations. Below we will show that there are also
differences in metallicities and kinematics.

%

\subsection{Spectroscopy}

\begin{figure*}
\centering
\includegraphics[width=8cm]{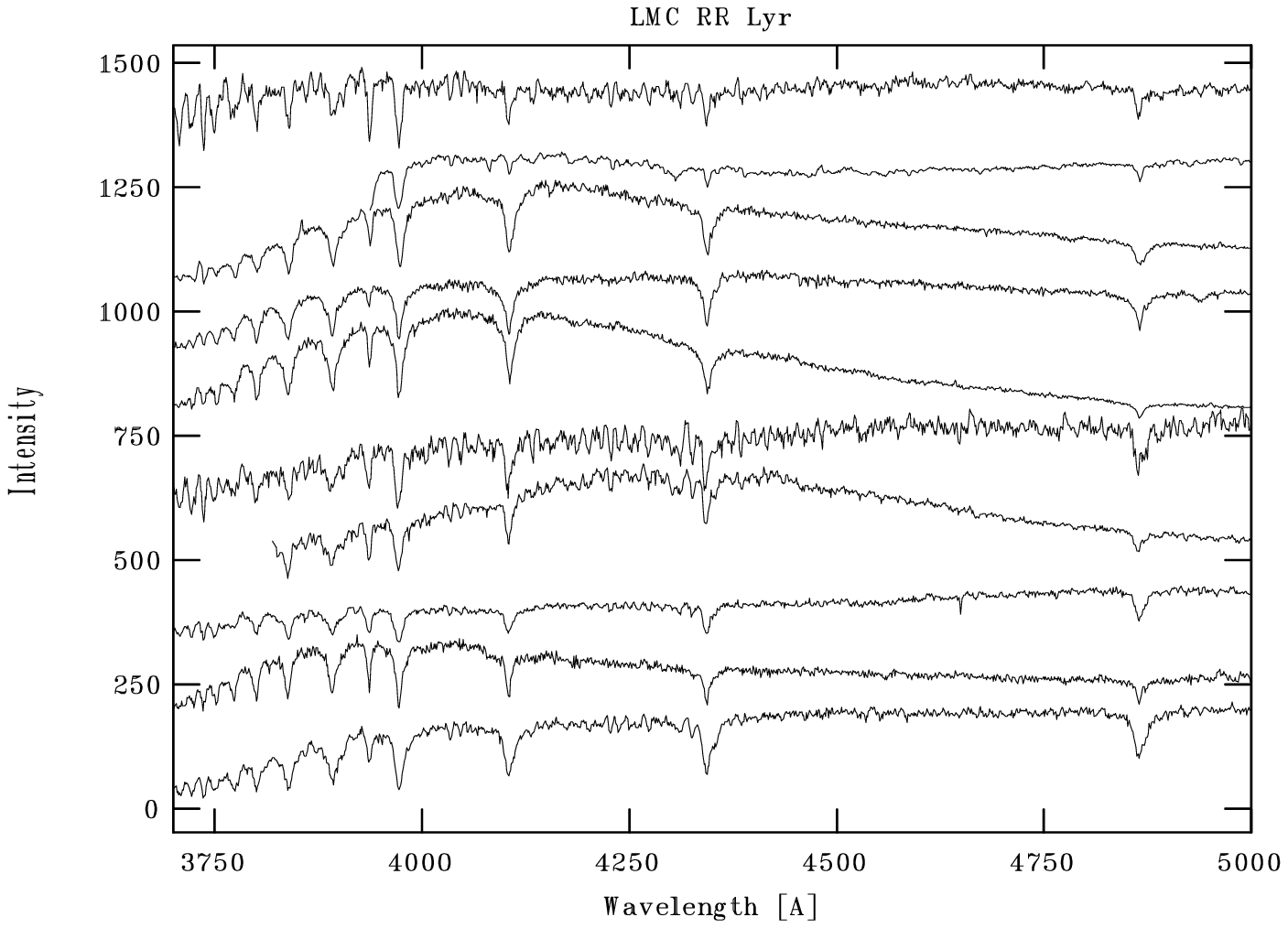}
\includegraphics[width=8cm]{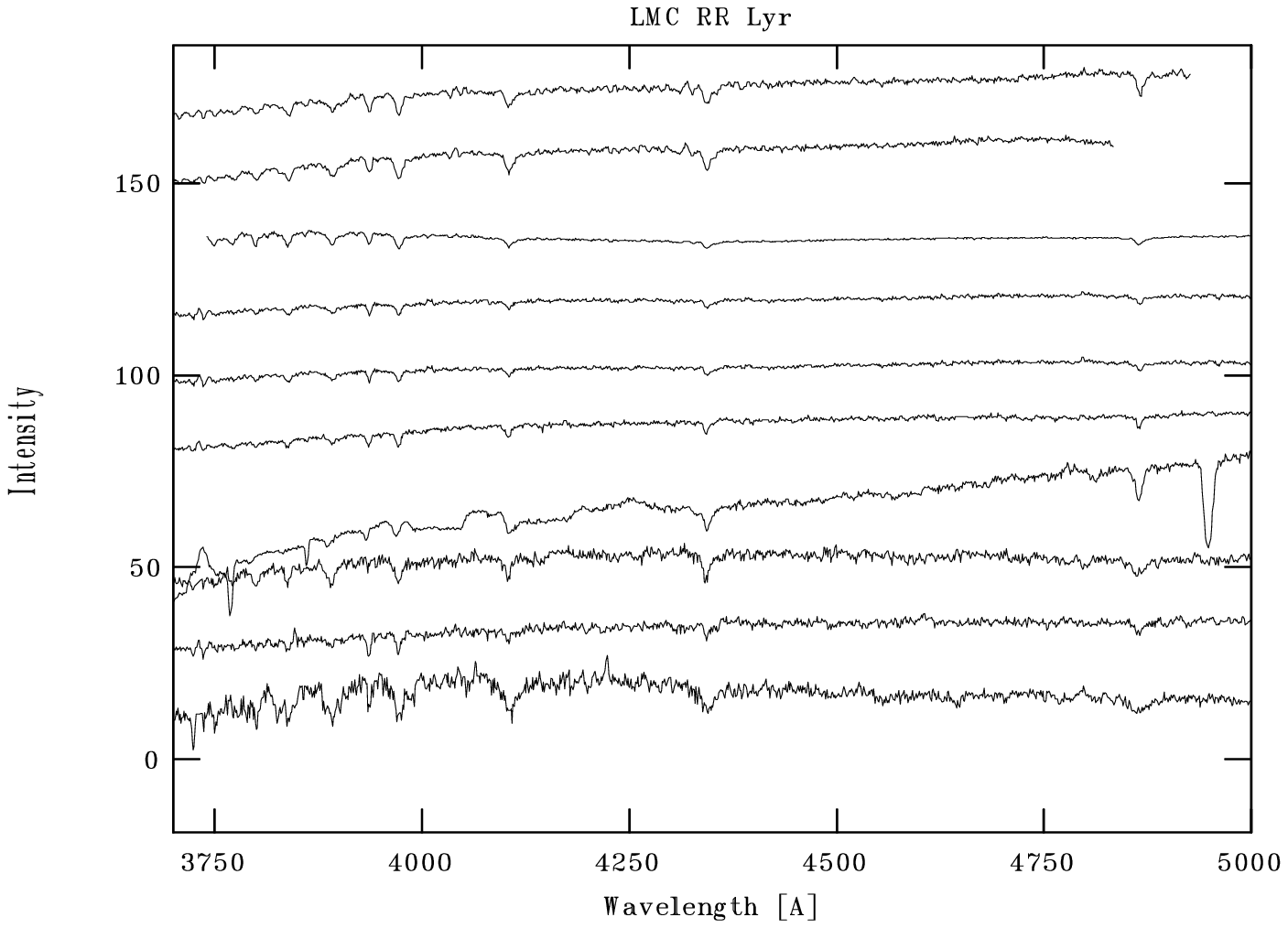}\\
\includegraphics[width=8cm]{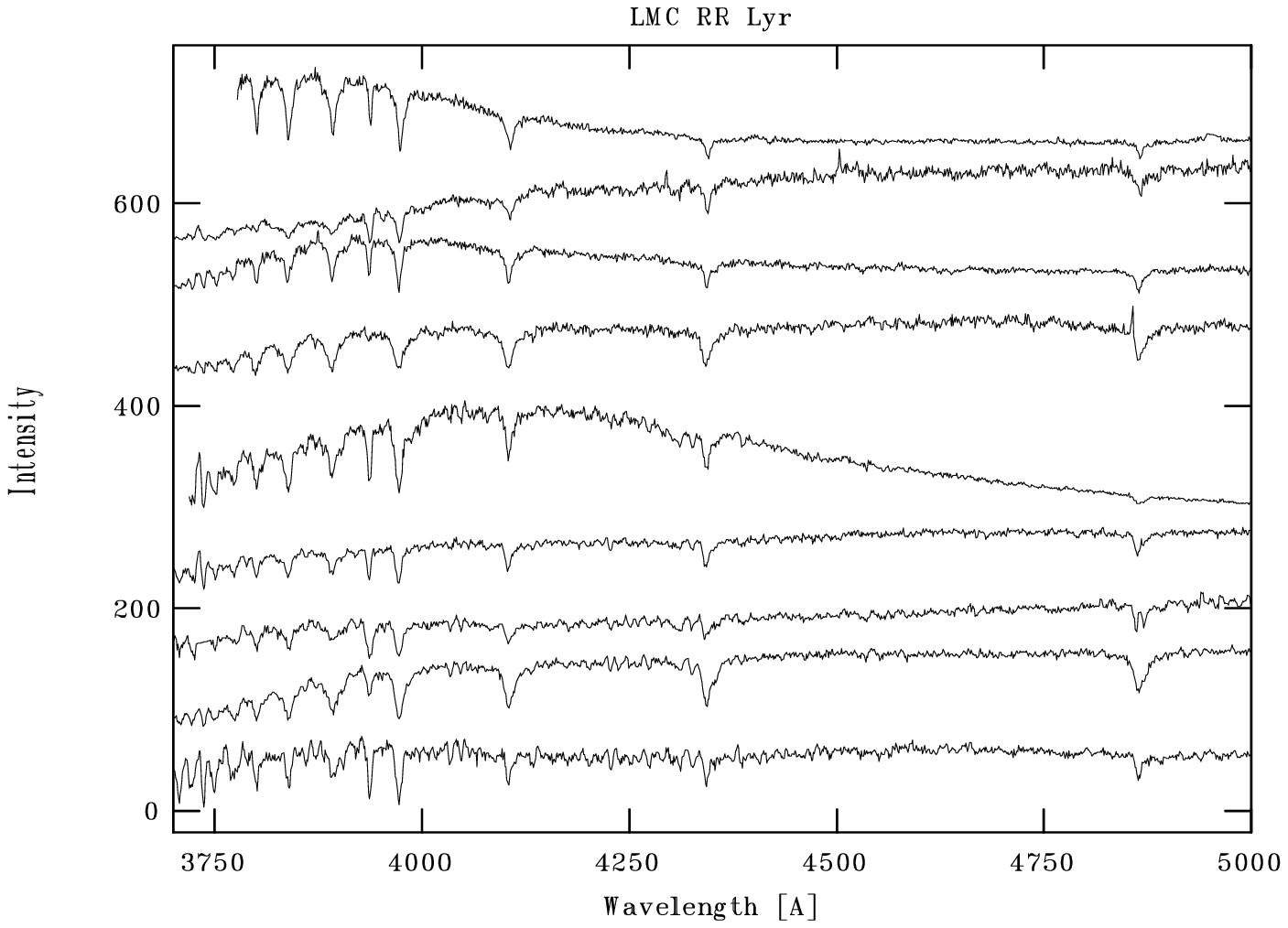}
\includegraphics[width=8cm]{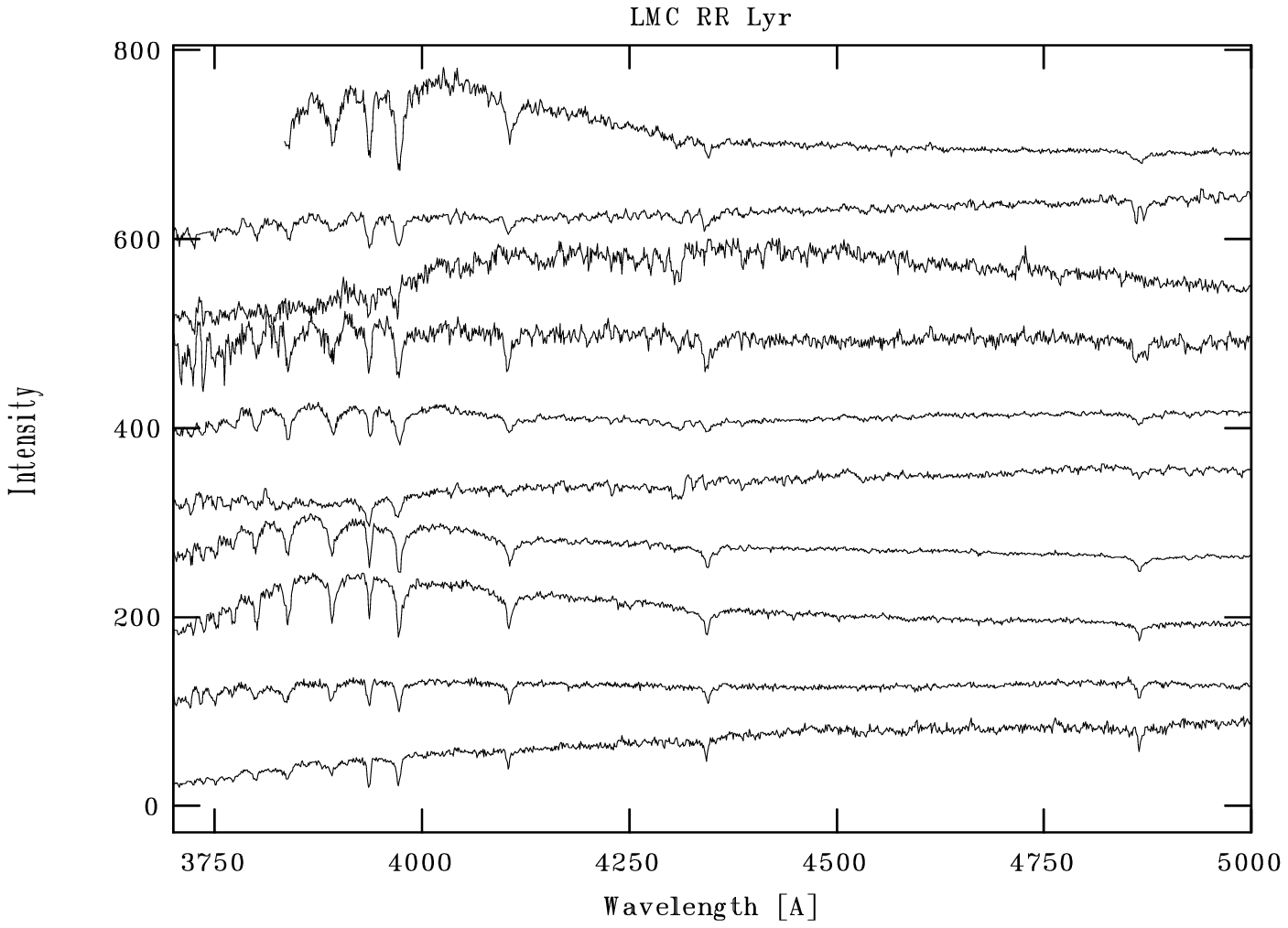}\\
  \caption[]{ Spectra of LMC RR Lyrae variables
	obtained with FORS1 in Jan. 2003.}
\label{RRLyrspectra}
\end{figure*}

\begin{figure}[h]
\resizebox{\hsize}{!}{\includegraphics{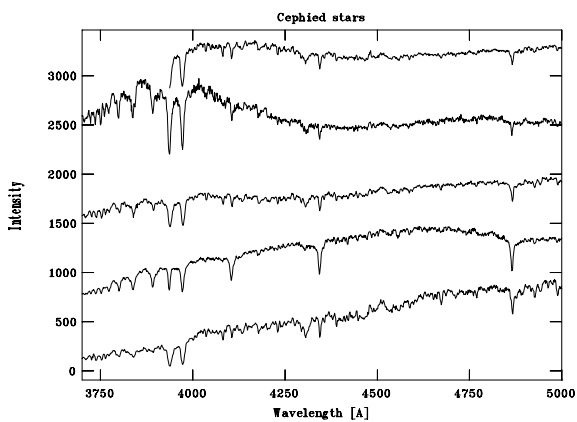}}
 \caption{ Spectra of Cepheid stars
obtained with FORS1 in Jan. 2003.
}
 \label{Fig04}
\end{figure}

\begin{figure}[h]
\resizebox{\hsize}{!}{\includegraphics{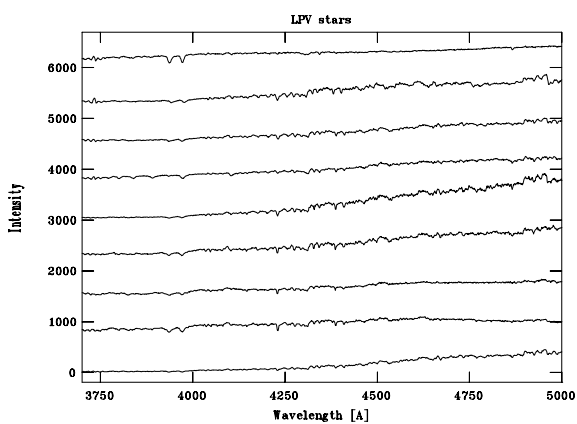}}
 \caption{ Spectra of OGLE LPV variables
obtained with FORS1 in Jan. 2003.
}
 \label{Fig05}
\end{figure}

The spectroscopic observations were taken with the FORS1 multi-slit 
spectrograph at the ESO Very Large Telescope (VLT) Unit Telescope 1 (UT1),
during the nights of 10 and 11 January 2003.
We used the GRIS\_600B+12 grating, that gives $R=1000$ and
covers from $\lambda 3500$ to $\lambda 5900$\AA.
This resolution is adequate for the measurement of radial velocities
even in the broad-lined RR Lyrae spectra, 
provided that a good S/N is achieved.
In  total, two exposures of 20 minutes were obtained  for each mask
containing 5-10 RR Lyrae stars.
As the FORS1 multi-slit spectrograph can take spectra of up to 19 objects
simultaneously, we placed LPV 
and Cepheid variables on the remaining slits. 
We observed 58 RR Lyrae,  5 Cepheids, and 23 Miras of the LMC.
The spectra of RR Lyrae variables are shown in Fig.~\ref{RRLyrspectra}. 
The spectra of Cepheid stars and LPV's are shown in  Fig.~\ref{Fig04} 
and Fig.~\ref{Fig05}.

%

%

\begin{figure}
\resizebox{\hsize}{!}{\includegraphics{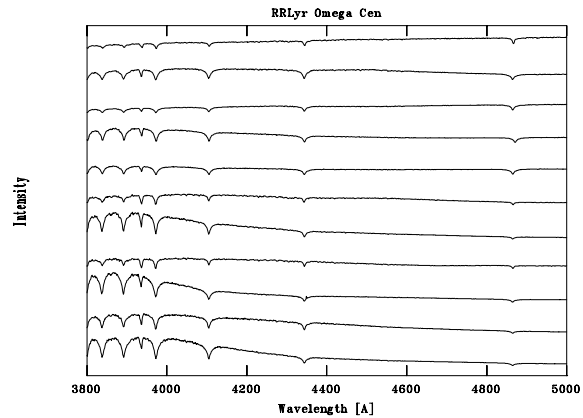}}
 \caption{Spectra of  $\omega$ Cen RR Lyrae variables
obtained with FORS1 in Jan. 2003.
}
 \label{Fig06}
\end{figure}

A substantial fraction of the observing time was devoted to calibrations.
Two masks containing RR Lyrae of the globular cluster $\omega$ Cen 
(Clement et al. 2001, Mayor et al. 1997,  Kaluzny et al. 1997) were
acquired using the same setup. Thus, high quality spectra of
17 $\omega$ Cen RR Lyrae (6 RRab, 7 RRc, and 4 RRe) were secured.
In addition, a few repeat observations of some of the
RR Lyrae stars were taken, in order to
assess the velocity errors.
The spectra of $\omega$ Cen 
RR~Lyrae stars are presented in Fig.~\ref{Fig06}.

The spectral data were 
reduced using  the standard  packages within IRAF.
HeNeAr lamps were used for the wavelength calibration, which typically
have 14 usable lines that yield 0.2 \AA~rms.
The final extracted and calibrated spectra have
on average S/N = 15. 
This is adequate to measure individual velocities good to about $10-30$ 
km s$^{-1}$.

\subsection{Photometry}

The $K$-band dataset  was obtained with
the SOFI infrared imager
at the European Southern Observatory's New Technology Telescope;
SOFI has a 1024$\times$1024 array with a pixel
size of 0.292 arcsec, with a total field
of view $5\times5$ arcmin.  
We obtained  two measurements for each star, 
in order to define the mean magnitude.
The photometry was derived with DAOPHOT II (Stetson 1994),
and calibrated with observations of 6 standard stars (Persson et al.~1998).
The 1-$\sigma$ standard deviation of the calibration solution
is 0.028~mag. The V and I bands photometry was obtained
with WFPC2 on board the HST (P.I. Cook, Alves at al. 2002). 
The HST photometry is single epoch F555W and F814W WFPC2 observations. 
Photometry with DAOPHOT II and the standard calibrations to the ground 
based system result in uncertainties of the order of 0.02-0.03 mag. 
The details of the data
reduction, error and completeness analysis of the IR and HST photometry
are given in Alves at al. (2002).

\section{Color - magnitude diagrams}

\begin{figure}[h]
\resizebox{\hsize}{!}{\includegraphics{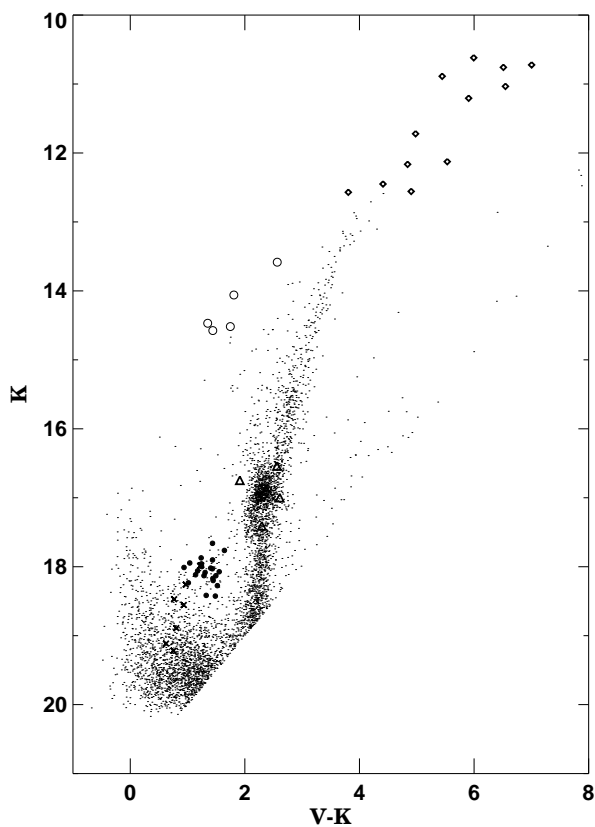}}
 \caption{ The 
	$K$,$(V-K)$ color-magnitude diagram of LMC fields 
	with MACHO variables overploted. 
The open circles are Cepheid stars, LPV stars are shown as squares, 
solid circles and crosses indicate RRab and RRc stars and triangles are for
RGB+RR Lyrae blends.
}
 \label{Fig10}
\end{figure}

Fig.~\ref{Fig10} shows the 
$K$,$(V-K)$ color-magnitude diagram (CMD). 
Well populated red giant branch (RGB) up to
$K=12.5$ and $V-K=4.2$ and 
red clump (RC) at $K=17$ are clearly visible. 
The variables from our sample, which have K measurements are superimposed: 
the open circles mark Cepheid stars, the LPV stars are shown as squares, 
solid circles indicated RRab Lyrae stars, crosses are for RRc stars
and triangles are for Red Giant Branch stars + RR Lyrae (RGB+RR Lyr) blends. 
Everywhere in this paper we use optical mean V, I and R magnitudes of variable stars,
taken from the MACHO and OGLE databases. 
The LPV stars are located at the tip of the giant branch,
RR Lyrae  and Cepheid stars are located within the Cepheid instability strips.
 Because of the difference between
FORS1 multi-slit spectrograph (6.8$\times$ 6.8 arcmin) 
and SOFI NTT fields (5$\times$ 5 arcmin)
only part of the stars with spectra have K magnitudes.
In total we have the K band magnitudes for 38 RR Lyrae stars, 
5 Cepheids and 12 LPV stars from MACHO database. 
Since we are using V and I mean magnitudes of the variable stars taken from 
MACHO and OGLE databases there is possibly a systematic 
error of up to ~0.10 mag, because
of the differences between MACHO and OGLE V and I mean magnitudes of
the variable stars and HST V and I photometry.

%

\begin{figure}[h]
\resizebox{\hsize}{!}{\includegraphics{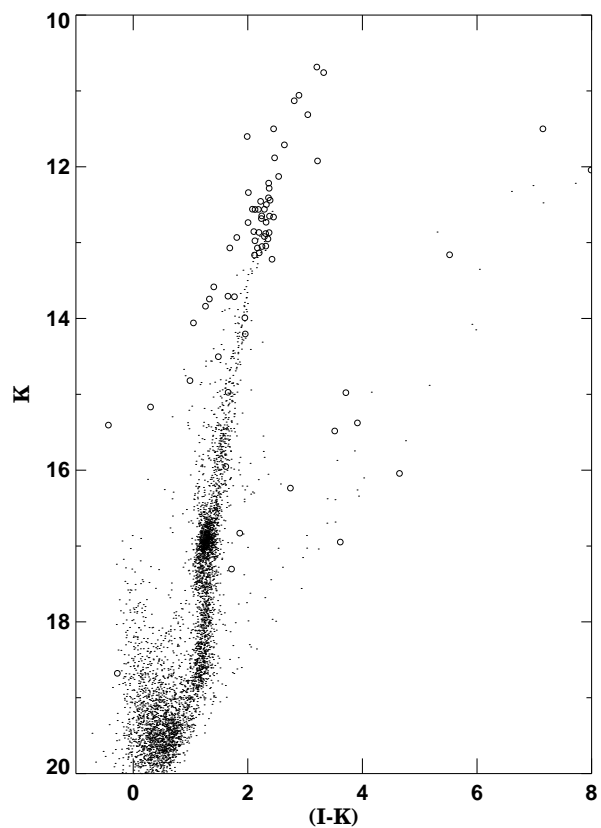}}
 \caption{The $K$,$(I-K)$ color-magnitude diagram of LMC fields with OGLE 
variables overploted.}
 \label{Fig11}
\end{figure}

Cross-identification with the OGLE database yields 70 additional variables
in these fields.
On the $(K,I-K)$  color-magnitude diagram (Fig.~\ref{Fig11}) 
they are overploted with large open circles. As can be seen most of them are
bright LPV stars. Unfortunately, in OGLE database we could not find accurate
classifications, periods or amplitudes of these stars.

%

Several OGLE variable stars lie well below the LMC  
red giant branch on Fig.~\ref{Fig11}. 
They are at the limit of the I-band photometry, and thus 
probably mismatched variables or noisy measurements.

\subsection{Boundaries of the instability strip}

The knowledge of the precise location of the RR Lyrae zone is a good test
for the stellar pulsation theory. Using the calibration of Montegriffo et al. (1998) we
calculated the effective temperature and bolometric magnitude of RR Lyrae stars
which have $(V-K)$ colors and K magnitudes. We assume E(B-V)=0.11 (Clementini et al. 2003).
A plot of ($K_{0}, (V-K)_{0}$) and ({$M_{bol}$, Log $T_{eff}$) of these variables
is show on Fig.~\ref{Fig12}. Open circles are for RRab stars, crosses stand for RRc stars. 
The theoretical value of the blue edge of the RR Lyrae instability strip 
is located near the $T_{eff}=7400K$
(Smith 1995), 
while the red edge is near
6100 $K$ (dashed lines on Fig.~\ref{Fig12}). Our calculations are in good agreement 
with the limit for the red edge: we calculated $T_{eff}=6150K$. For the blue edge, however
we have five stars which are hotter, than the theoretical blue limit. Careful check of these
stars show that three of them could be anomalous - one is an RRab with large period:
P=0.616 days and relatively small amplitude: $A_{V}=0.369$ mag, 
two are RRc stars with larger than usual amplitudes in both V and 
R: $A_{V}=A_{R}=0.78$.  Taking into consideration
the errors of our $V-K$ colors and calibration to the theoretical plane, we estimate
the error of our $T_{eff}$ determination to be approx. 100K. 
Since our mean K-band magnitudes are based on two random
measurements, this can add additional uncertainty of appox. 0.15 in colors or 250K
in temperature. Thus we calculate for the blue edge $7550\pm270K$.

\begin{figure}[h]
\resizebox{\hsize}{!}{\includegraphics{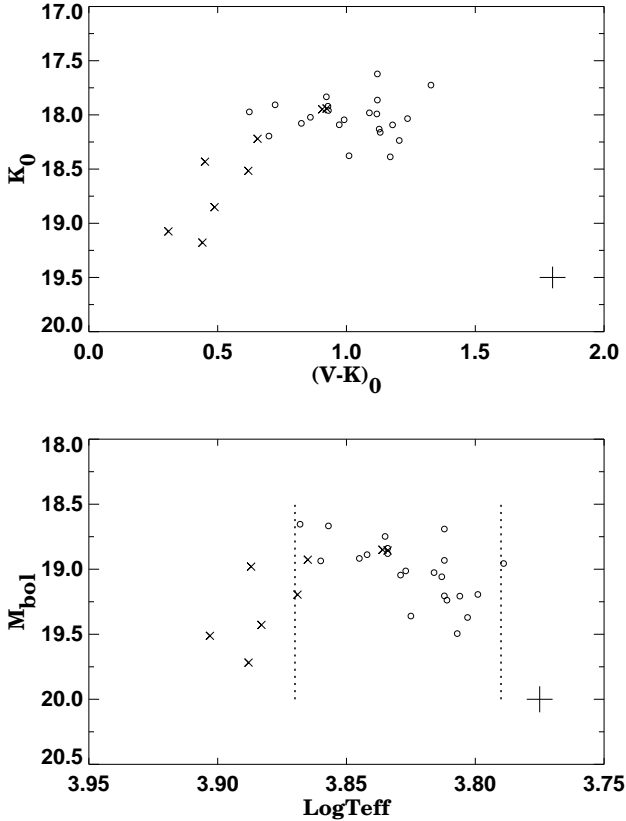}}
\caption{Top panel:  
A plot of $(V-K)_{0}$ vs. $K_{0}$ of the RR Lyrae stars
with K-band magnitudes. Open circles are for RRab stars, crosses stand for RRc stars.
Bottom panel: A plot of Log $T_{eff}$ vs. $M_{bol}$. The dashed lines represent theoretical
limits of the blue and red edges as given in Smith (1995). The crosses represent
the typical errors of the photometry and transformation to the theoretical plane.}
\label{Fig12}
\end{figure}

\subsection{ Period-Amplitude  relations}

The period--amplitude diagram, known as the Bailey diagram, is a widely used tool
for analyzing features of RR Lyrae stars. Empirical and theoretical studies
suggest that the distribution of RR~Lyrae stars 
in the Bailey diagram depends on their metallicity. 
The period - $A_{V}$ (from MACHO databases) amplitude diagram
of the RR Lyrae stars in our fields is 
shown in the left panel of Fig.~\ref{Fig13}. 
Since we have only two K images per fields, 
it was not possible to determine the 
K band amplitudes. In general, there is a  
clear separation between RRab (open circles), 
RRc (crosses) and RRe (filled squares). 
Fundamental mode RR Lyrae stars present an anti-correlation between the period
and amplitude, which seems to be linear. 
Soszynski et al. (2003) found non-linearity in these diagrams 
using approx. 7600 RR Lyrae stars in the LMC. Our result 
is probably due to our small sample (only 70 stars).
The width of the sequence is believed to
be an effect of a spread in the metal content, which means that our
sample has a relatively small metallicity spread.
For RRc stars, we also found a weak anti-correlation of amplitudes
and periods (Soszynski et al. 2003). 

In the middle and right panels of 
Fig.~\ref{Fig13} the (V-K) color - LogP 
and the (V-K) color - $A_{V}$ amplitude diagrams are shown. 
In general, these relations 
follow the same trends as amplitudes, 
although they are not so clearly visible, forming
overlapping sequences. 

According to the theoretical predictions of Bono et al. (1997) the
luminosity amplitude depends marginally on the metallicity up to
[Fe/H]=0.7 dex. For more metal-rich stars an increase in the metallicity
causes a decrease in the amplitudes.

\begin{figure}[h]
\resizebox{\hsize}{!}{\includegraphics{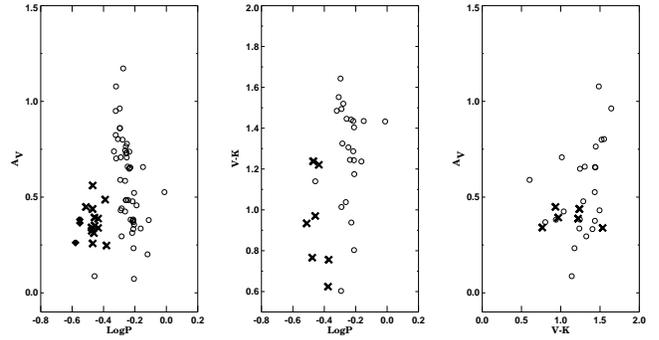}}
\caption{Left panel:
The relation between V amplitudes and Log P for RR Lyrae
stars. Open circles are for RRab stars, crosses for RRc stars
and filled squares for RRe stars. 
Middle and right panels: The relations between $V-K$ color and Log P
and $A_{V}$. The symbols are the same as above.}
\label{Fig13}
\end{figure}

Our sample was derived from a preliminary classification of MACHO
variables.  Subsequent analysis has shown MACHO star
11.8750.1425 is not an RRab, while MACHO star 80.648.3667 is an RRab
(Alcock et al. 2003). 
MACHO star 2.5507.6257 has not been confirmed to be an RRc, nor has
MACHO star 11.8750.1425.
These stars are omitted from the plots and calculations in the following
sections.

\subsection{ Period-luminosity  relations}
 
The average apparent luminosities (coming from MACHO databases)
of 70 RR Lyrae stars is $<V> = 19.45\pm0.04$ and average 
K luminosities of 33 RR Lyrae stars is $<K>=18.20\pm0.06$.
The RGB+RR Lyrae blends are not used for this calculation. 
The average MACHO database V luminosity of our 
sample is in agreement with Clementini et al (2003)
determination ($<V> = 19.412$ ) with a systematic shift of 0.038. 
Fig.~\ref{Fig14} shows $V_0$ vs. LogP and  K-band vs. LogP, with
magnitudes corrected for reddening.
RRab Lyrae star are shown with solid circles, 
the RRc stars with open ones. Since the sample of RRc stars is limited, their periods
are fundamentalized.
There is no obvious relation 
between $V_0$ magnitude and LogP, 
while the RR Lyrae K band magnitudes show a well defined linear trend.
Bono et al. (2001) published a new theoretical 
near-infrared Period-Luminosity-Metallicity relation and 
more recently it was improved on the basis of up-to-date pulsating models (Bono et al. 2003).
The predicted theoretical relation for fundamental mode pulsators 
(solid line) is overploted on Fig.~\ref{Fig14}. 
The dashed line represents the best fit for RRab and fundamentalized RRc stars obtained
using all data, the thin line is the same fit after the three faint outliers and RRab star
with very large period were excluded.
The agreement 
with Bono et al. (2003) prediction is good, when we exclude these four
outliners. Following the  same
procedure as in Bono et al. (2001) we calculated $M_K$ = 0.332 
at LogP= -0.30, obtaining an infrared distance modulus of $\mu_{0}(K) = 18.48\pm0.08$.
  Here, we adopted $E(B-V) = 0.11$ (Clementini et al 2003) 
and $\mathrm{[Fe/H]} = -1.46$  dex (see Sec~8) and luminosity 
$\log (L/L\sun) = 1.72$. 
The calculated distance is comparable within the errors 
with most of the LMC distance determinations
from the literature.

%
\begin{figure}[h]
\resizebox{\hsize}{!}{\includegraphics{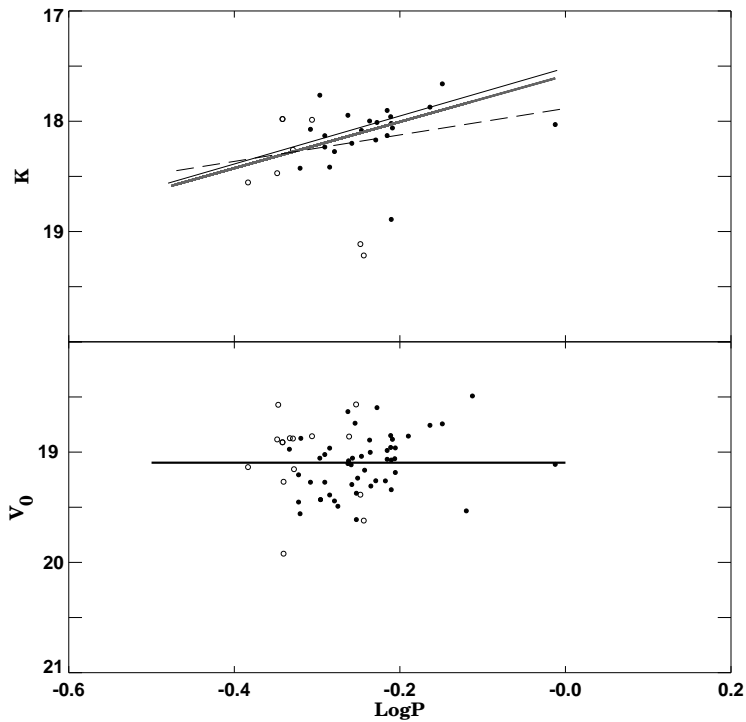}}
\vspace{0.5cm}
 \caption{Top panel: LogP vs. K plot for RRab stars (sold circles) 
	and RRc (open circles) in the  LMC. The Bono et al. (2003) relations 
        are shown with solid lines. The dashed line is a best fit for RRab and
        fundamentalized RRc stars, the thin line represent the best fit, without the
        four outliers.
	Bottom panel: LogP vs. $V_0$ plot for RRab (solid circles) and fundametalized 
	RRc (open circles) in the LMC. The solid line is  the average magnitude.
	The adopted distance module, reddening and metallicity 
	are $18.48$, $0.11$ and $-1.48$, respectively. 
}
 \label{Fig14}
\end{figure}

\section{Radial Velocities}

It is well known (Smith 1995) that the line spectrum of 
RR Lyr star undergoes a cyclic change during the pulsation cycle. 
The observed radial velocity depends also upon which spectral
lines are measured. In some phases of the pulsation cycle, the velocities
based on measurements of the Balmer lines are systematically different 
from those based on metallic lines. The reason is that lines of
different elements and ionization states arise from different levels within the
moving stellar atmosphere. The difference is largest during rising light.
The difference in the radial velocities of lines that form at high
and low optical depths is due to a radial velocity gradient. This
phenomenon also causes a systematic shift in the phase of maximum
velocity (Bono et al. 1994).

There are two commonly used techniques to measure radial velocities:
cross-correlation technique, which computes the radial velocities via
Fourier cross correlation, and by centroiding the individual spectral 
lines. We measured the radial velocities of our sample using both of
these techniques.
The comparison between radial velocities measured by two methods 
for $\omega$ Cen RR Lyrae 
stars is shown on Fig.~\ref{Fig15}. The cross-correlation velocities 
are corrected for heliocentric velocity and are shifted to the mean 
velocity of $\omega$ Cen $RV_{mean}=232.78\pm13.7$ km s$^{-1}$ (Mayor  at al. 1997).
In general, the agreement is good, 
except for a few cases with larger differences.
We calculated the phases of all RR Lyrae stars at the moment of
our observations using  ephemeredes, amplitudes and periods from Kaluzny et al. 
(1997). As expected, these outlying stars (V266, V88, V271, V275 and V267 see Table~2) 
are around maximum light and the radial velocities of
CaII-lines are significantly different from those of H-lines.
In general, the two techniques yield similar results within 
15 km s$^{-1}$, but the individual
errors of measurements are larger when we use the cross-correlation technique.
The result is not surprising, taking into account the relatively broad lines 
of RR Lyrae stars.

%
\begin{figure}[h]
\resizebox{\hsize}{!}{\includegraphics{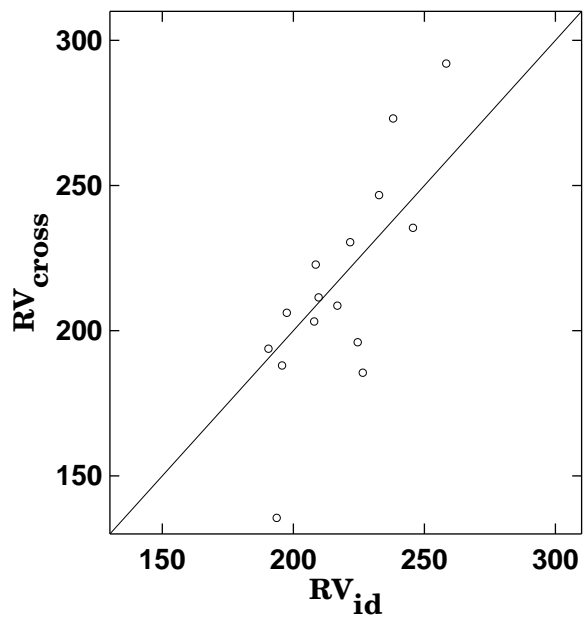}}
 \caption{Radial velocities in km s$^{-1}$ of  $\omega$ Cen RR Lyrae variables
measured by Fourier cross correlation and by centroiding the
individual spectral 
lines techniques.
}
 \label{Fig15}
\end{figure}

Therefore, we prefer to measure the radial velocities as an unweighted
mean value of individual radial velocities of H$\beta$,
H$\gamma$, H$\delta$, and CaII K $\lambda$3933.66 \AA, excluding the Ca line
velocities in the cases of rising light. 
We measured radial velocities of 58 RR Lyrae stars, 5 Cepheids and 33 LPV's 
stars in our six LMC fields. The internal errors measured from the different
lines range from 1 to 33 km s$^{-1}$.

The pulsation of a 
RR Lyrae star induces a periodic variation in the radial velocity of the 
star. The radial velocity curve roughly reflects the light curve, with
minimum radial velocity coming at maximum visible light 
for RRab variables.
For RRc stars the maximum radial velocity is reached about 0.1 of a cycle
after maximum light. The velocity  amplitudes for RRab stars are
between 60-70 km s$^{-1}$. 
and for RRc stars are between 30-40 km s$^{-1}$ (Smith 1995). 
Therefore, it is necessary to correct such measured radial velocities 
for this large velocity amplitude through their cycle of pulsation. 
Liu  (1991) derived a correlation between the pulsational velocity 
vs. V light amplitudes for RRab stars.
Unfortunately, we have only 6 RRab stars in our control sample of $\omega$ Cen
RR Lyrae's and 
two of them have only one line measured accurately. We used the Liu  (1991)
calibration to correct the raw velocities of these four 
stars and  to estimate how
the  velocity dispersion will change after the correction.
The mean value of measured radial velocities is $RV_{mean}=247\pm29$ km s$^{-1}$.
The pulsation amplitudes are between 67 and 83 km s$^{-1}$ and the mean value
of the corrected velocities is  $RV_{mean}=264\pm19$ km s$^{-1}$.  
For RRc stars Liu (1991) says that there should be a similar relation, but he
had too few data to derive it.
Although the pulsation amplitudes are large since the RR Lyrae are
observed in random phase, the velocity dispersion does not change so much.
The data for $\omega$ Cen stars are summarized in Table~2.

\begin{table}[t]\tabcolsep=0.5pt\small
\begin{center}
\caption{$\omega$ Cen RR Lyrae stars}
\label{Table2}
\begin{tabular}{l@{ }c@{ }c@{}c@{ }c@{ }l@{ }c@{}l@{ }c@{ }c@{ }}
\hline
\multicolumn{1}{c}{Name}&
\multicolumn{1}{c}{OGLE}&
\multicolumn{1}{c}{$<V>$}&
\multicolumn{1}{c}{$A_{V}$\hspace{0.2cm}}&
\multicolumn{1}{c}{Period}&
\multicolumn{1}{c}{Type}&
\multicolumn{1}{c}{$RV$\hspace{0.2cm}}&
\multicolumn{1}{c}{$\sigma_{RV}$}&
\multicolumn{1}{c}{Phase}&
\multicolumn{1}{c}{$RV_{corr}$}\\
\hline
V139&	118&	14.35&	0.71&	0.6768&	RRab&	278&\hspace{0.2cm}	 4& 	 	0.09&		282\\
V111&	136&	14.46&	0.60&	0.7629&	RRab&	228&\hspace{0.2cm}	 16&	 	0.16&		238\\
V118&	109&	14.43&	1.00&	0.6116&	RRab&	218&\hspace{0.2cm}	 27&	 	0.54&		263\\
V88&	210&	14.23&	0.67&	0.6904&	RRab&	266&\hspace{0.2cm}	 13&            0.97&  	        273\\
V144&	112&    14.41& 	0.48&   0.8352& RRab&   239&\hspace{0.2cm}       $-$&           0.75&           275\\
V113&   128&    14.41& 	1.20&   0.5733&	RRab&   148&\hspace{0.2cm}       $-$&           0.76&           209\\       
V273&	139&	14.60&	0.39&	0.3672&	RRc&	237&\hspace{0.2cm}	 3&	        0.09&		$-$\\		 
V153&	140&	14.55&	0.44&	0.3863&	RRc&	245&\hspace{0.2cm}	 15&	 	0.24&		$-$\\		 
V264&	108&	14.75&	0.45&	0.3214&	RRc&	211&\hspace{0.2cm}	 5&	        0.26&		$-$\\		 
V145&	130&	14.56&	0.44&	0.3732&	RRc&	230&\hspace{0.2cm}       10&       	0.36&		$-$\\		 
V157&	138&	14.56&	0.45&	0.4066&	RRc&	247&\hspace{0.2cm}	 13&	 	0.66&		$-$\\		 
V275&	149&	14.50&	0.34&	0.3782&	RRc&	214&\hspace{0.2cm}       17&	 	0.80&		$-$\\		 
V271&	132&	14.44&	0.43&	0.4431&	RRc&	229&\hspace{0.2cm}	 12&	 	0.97&		$-$\\		 
V119&	106&	14.65&	0.29&	0.3059&	RRe&	216&\hspace{0.2cm}	 2& 	 	0.26&		$-$\\		 
V166&	142&	14.54&	0.13&	0.3414&	RRe&	242&\hspace{0.2cm}	 8&	        0.51&		$-$\\		 
V267&	121&	14.46&	0.25&	0.3159&	RRe&	253&\hspace{0.2cm}	 3& 	 	0.78&		$-$\\		 
V266&	115&	14.54&	0.27&	0.3524&	RRe&	258&\hspace{0.2cm}	 6& 	 	0.86&		$-$\\		
\hline
\end{tabular}
\end{center}
\end{table}

The same test was performed on the sample of RRab stars in the LMC.
We used the MACHO database to derive the phases of the stars in the moment
of our observations. The mean value of measured radial velocities is $RV_{mean}=228\pm61 $ km s$^{-1}$,
the mean value of the corrected velocities is  $RV_{mean}=252\pm58$ km s$^{-1}$. 
The difference in the velocity dispersions for corrected and un-corrected radial
velocities is small.
 
We decided to use un-corrected radial velocities for the 
calculation of the velocity dispersion for the whole sample, 
rather than correcting only part of the sample. The comparison of the 
velocity dispersion of RR Lyrae in the LMC with the other 
stellar populations will then be used to test for the presence of a stellar
halo around the galaxy.

The data for the 
LMC RR Lyrae stars and Cepheids are summarized in Table~3, 
and in Table~4 we list the radial velocities of OGLE LPVs.

In columns 2,3,4 and 5 of Table~3 are mean V magnitude, amplitude,
period of the pulsation and type of the star taken from
MACHO database. Note uncertain/wrong 
classification of 11.8750.1425,80.6468.3667 and 2.5507.6257 (see Sec 3.1 and 3.2).

\begin{table}[t]\tabcolsep=0.1pt\tiny
\begin{center}
\caption{LMC RR Lyrae and Cepheid stars}
\label{Table3}
\begin{tabular}{l@{ }c@{ }c@{}c@{ }l@{ }c@{ }l@{ }}
\hline
\multicolumn{1}{c}{MACHO name}&
\multicolumn{1}{c}{$<V>$}&
\multicolumn{1}{c}{$A_{V}$}&
\multicolumn{1}{c}{Period(d)}&
\multicolumn{1}{c}{Type}&
\multicolumn{1}{c}{$RV$ (km/s)}&
\multicolumn{1}{c}{$\sigma_{RV}$}\\
\hline
10.3802.311		&	18.99	&	0.43	&	0.5460	&	RRab	&	182	&	10	\\
10.3802.339		&	19.37	&	0.71	&	0.5120	&	RRab	&	173	&	6	\\
10.3802.446		&	19.49	&	0.45	&	0.3080	&	RRc	&	82	&	13	\\
10.3922.978		&	19.09	&	0.48	&	0.5568	&	RRab	&	315	&	$-$	\\
10.3923.351		&	19.24	&	0.34	&	0.3340	&	RRc	&	144	&	19	\\
11.8622.757		&	19.11	&	0.34	&	0.6860	&	RRab	&	178	&	18	\\
11.8623.3792	&	19.31	&	0.38	&	0.6149	&	RRab	&	187	&	11	\\
11.8623.779		&	19.20	&	0.38	&	0.6150	&	RRab	&	196	&	11	\\
11.8623.826		&	18.95	&	0.38	&	0.5902	&	RRab	&	288	&	20	\\
11.8743.1422	&	20.27	&	0.56	&	0.3396	&	RRc	&	179	&	11	\\
11.8744.658		&	18.92	&	0.25	&	0.4155	&	RRc	&	312	&	8	\\
11.8744.752		&	19.14	&	0.37	&	0.2809	&	RRe	&	210	&	9	\\
11.8744.830		&	19.47	&	0.48	&	0.5510	&	RRab	&	330	&	19	\\
11.8749.1208	&	19.56	&	0.82	&	0.4757	&	RRab	&	257	&	23	\\
11.8749.1324	&	19.63	&	0.43	&	0.5120	&	RRab	&	282	&	$-$	\\
11.8750.1425	&	19.26	&	0.09	&	0.3490	&	RRab	&	270	&	28	\\
11.8750.1672	&	19.62	&	0.26	&	0.3400	&	RRc	&	231	&	14	\\
11.8750.1827	&	19.74	&	0.30	&	0.5186	&	RRab	&	233	&	11	\\
11.8750.2045	&	19.81	&	0.95	&	0.4763	&	RRab	&	267	&	33	\\
11.8870.1275	&	19.33	&	0.74	&	0.4638	&	RRab	&	247	&	19	\\
11.8871.1096	&	19.43	&	0.74	&	0.5473	&	RRab	&	291	&	$-$	\\
11.8871.1122	&	19.71	&	0.70	&	0.5010	&	RGB+RR	&	294	&	$-$	\\
11.8871.1299	&	19.61	&	0.56	&	0.5960	&	RGB+RR	&	233	&	22	\\
11.8871.1362	&	19.61	&	0.31	&	0.6064	&	RRab	&	339	&	8	\\
11.8871.1447	&	19.59	&	0.26	&	0.2640	&	RRe	&	148	&	6	\\
11.8871.1516	&	19.46	&	0.59	&	0.5463	&	RRab	&	309	&	$-$	\\
13.5839.1023	&	19.66	&	0.65	&	0.5823	&	RRab	&	219	&	$-$	\\
13.5840.608		&	19.32	&	0.44	&	0.5188	&	RRab	&	234	&	14	\\
13.5840.730		&	19.41	&	0.07	&	0.6215	&	RRab	&	228	&	24	\\
13.5960.884		&	19.23	&	0.31	&	0.3459	&	RRc	&	183	&	11	\\
13.5962.547		&	19.21	&	0.46	&	0.6464	&	RRab	&	207	&	14	\\
13.6082.701		&	19.41	&	0.73	&	0.5525	&	RRab	&	214	&	8	\\
2.5507.5945		&	19.54	&	0.52	&	0.6234	&	RRab	&	256	&	10	\\	
2.5507.6046		&	19.63	&	0.80	&	0.4920	&	RRab	&	338	&	31	\\	
2.5508.3096		&	19.96	&	0.71	&	0.5593	&	RRab	&	149	&	29	\\	
2.5628.5690		&	19.69	&	0.37	&	0.6156	&	RRab	&	144	&	18	\\	
2.5628.6276		&	19.91	&	1.08	&	0.4781	&	RRab	&	135	&	$-$	\\	
79.5507.1039	&	19.59	&	0.78	&	0.5608	&	RRab	&	238	&	14	\\	
79.5507.1485	&	19.31	&	0.36	&	0.6234	&	RRab	&	261	&	9	\\	
79.5507.1580	&	19.46	&	0.53	&	0.9720	&	RRab	&	209	&	1	\\	
79.5508.427		&	19.73	&	0.73	&	0.5593	&	RRab	&	200	&	15	\\	
79.5508.534		&	19.52	&	0.48	&	0.5723	&	RRab	&	169	&	$-$	\\	
79.5508.682		&	19.89	&	0.20	&	0.7591	&	RRab	&	161	&	13	\\	
79.5508.735		&	19.78	&	0.86	&	0.5057	&	RRab	&	215	&	$-$	\\	 
79.5628.1065	&	19.84	&	1.17	&	0.5305	&	RRab	&	274	&	19	\\	
79.5628.1300	&	19.42	&	0.33	&	0.6160	&	RRab	&	280	&	$-$	\\	
79.5628.1597	&	19.51	&	0.38	&	0.2812	&	RRe	&	179	&	3	\\	
79.5628.1650	&	19.26	&	0.44	&	0.3389	&	RRc	&	147	&	$-$	\\
79.5628.2110	&	19.80	&	0.80	&	0.5260	&	RRab	&	247	&	$-$	\\
80.6347.1940	&	19.21	&	0.49	&	0.4083	&	RRc	&	184	&	14	\\
80.6467.2128	&	19.35	&	0.74	&	0.5805	&	RRab	&	151	&	14	\\
80.6468.1883	&	18.92	&	0.32	&	0.3346	&	RRc	&	266	&	11	\\
80.6468.2616	&	19.21	&	0.39	&	0.3684	&	RRc	&	199	&	11	\\
80.6468.3667	&	20.05	&	1.19	&	0.4308	&	RRc	&	101	&	1	\\
80.6469.1657	&	19.24	&	0.65	&	0.5804	&	RRab	&	122	&	12	\\
80.6469.1712	&	18.84	&	0.38	&	0.7720	&	RRab	&	232	&	$-$	\\
80.6588.2703	&	19.23	&	0.70	&	0.4794	&	RRab	&	259	&	$-$	\\
80.6589.2425	&	19.51	&	0.35	&	0.3497	&	RRc	&	198	&	11	\\
80.6348.23		&	15.70	&	0.37	&	4.7348	&	Cep	&	328	&	$-$	\\		
80.6468.21		&	15.52	&	0.12	&	3.1557	&	Cep	&	285	&	6	\\		
80.6468.46		&	15.82	&	0.58	&	3.3587	&	Cep	&	278	&	6	\\		
11.8622.24		&	15.90	&	0.24	&	2.0921	&	Cep	&	241	&	34	\\		
11.8870.40		&	16.26	&	0.59	&	2.9827	&	Cep	&	236	&	10	\\		 
\hline
\end{tabular}
\end{center}
\end{table}
\clearpage

\begin{table}[t]\tabcolsep=0.1pt\tiny
\begin{center}
\caption{LPV stars from OGLE database}
\label{Table4}
\begin{tabular}{l@{ }c@{ }l@{ }}
\hline
\multicolumn{1}{c}{OGLE name}&
\multicolumn{1}{c}{$RV$ (km/s)}&
\multicolumn{1}{c}{$\sigma_{RV}$}\\
\hline
OGLE05145362-6846533	&	214	&	39 	\\
OGLE05145091-6850289	&	247	&	10 	\\
OGLE05145362-6846533	&	272	&	$-$	\\
OGLE05144133-6848591	&	261	&	17 	\\
OGLE05143819-6849131	&	223	&	40 	\\
OGLE0514363-6849378	&	269	&	27 	\\
OGLE05143421-685033	&	193	&	31	\\
OGLE05144922-684628	&	160	&	$-$	\\
OGLE05144133-6848591	&	238	&	1 	\\
OGLE05143815-6849217	&	226	&	34	\\
OGLE05143443-6849511	&	207	&	13 	\\
OGLE05143421-685033	&	218	&	29	\\
OGLE05041263-6931104	&	164	&	17	\\
OGLE0504633-6931282	&	203	&	49 	\\
OGLE05041042-6931483	&	193	&	34	\\
OGLE0504966-6932445	&	224	&	31	\\
OGLE05202161-6912358	&	81	&	$-$	\\
OGLE05201061-6913378	&	243	&	36 	\\
OGLE05202958-6914441	&	218	&	$-$	\\
OGLE05201862-6916558	&	344	&	2	\\
OGLE05202644-6913122	&	47	&	2	\\
OGLE0520061-6913378	&	143	&	78 	\\
OGLE05202958-6914441	&	253	&	32 	\\
OGLE05202789-691596	&	211	&	32 	\\
OGLE0520484-691531	&	201	&	37 	\\
OGLE05202923-6915589	&	277	&	50 	\\
OGLE05203023-6916347	&	206	&	4 	\\
OGLE05202042-6917356	&	399	&	$-$	\\
OGLE05344257-702531	&	261	&	26 	\\
OGLE05343669-7027283	&	151	&	$-$	\\
OGLE05343389-7027361	&	213	&	$-$	\\
OGLE05345159-7026341	&	275	&	$-$	\\
OGLE05344708-7025235	&	294	&	$-$	\\
OGLE0534446-7027311	&	174	&	35 	\\
\hline
\end{tabular}
\end{center}
\end{table}

We discarded 15 RR Lyrae stars that have only one 
line measured accurately, leaving  43 stars that have 2-4 lines accurately
measured. 
Of the final 43 stars considered, 29 are RRab, 11 are RRc and 3 are RRe
according to the MACHO light curves.
The same analysis was performed for LPV stars - here we have 23 stars with 
accurately measured velocities.

In the  Fig.~\ref{Fig16} the histogram of radial velocities of 
the whole sample of RR Lyrae and LPV
stars are plotted with the solid lines 
and the selected samples, with the best measured
velocities, are shown with the dotted lines.
As can be seen the distributions are similar.

%
\begin{figure}[h]
\resizebox{\hsize}{!}{\includegraphics{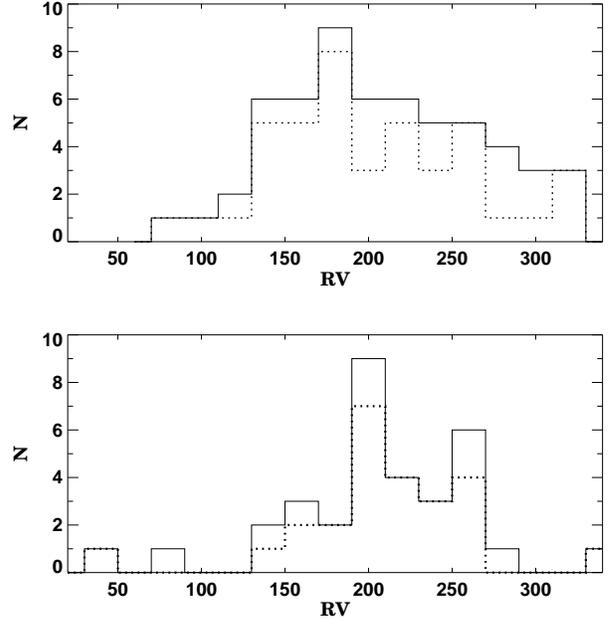}}
 \caption{ The radial velocity histogram for RR Lyrae stars (top panel) and 
	LPVs (bottom panel) in our fields. The dotted lines are for the selected sample of stars,
	 solid lines marked the whole sample (see text).}
 \label{Fig16}
\end{figure}

For the LMC RR Lyrae we measure $\sigma_{obs}=61$ km s$^{-1}$ and 
$\sigma_{true}= 53$ km s$^{-1}$ (Minniti et al. 2003). 
This is larger than the velocity dispersion of other populations: 
$\sigma_{true}= 28$ km s$^{-1}$ for 
the LMC LPV stars, $\sigma_{true}= 24$ km s$^{-1}$ 
for Cepheids and  $\sigma_{true}= 19$ km s$^{-1}$ for $\omega$ Cen RR Lyrae stars.
This large velocity dispersion of 
$\sigma_{true}= 53$ km s$^{-1}$
implies that RR Lyrae stars are distributed 
in a halo population as discussed in (Minniti et al. 2003).

In order to estimate the stability of the measured dispersion, we 
divide the sample in different sub samples. 
For the best N=29 RRab stars,  $V_{mean}=228$ km s$^{-1}$, $\sigma=60$ km s$^{-1}$; 
for all N=43 RRab stars,  $V_{mean}=233$ km s$^{-1}$, $\sigma=57$ km s$^{-1}$; 
for the best N=11 RRc stars,  $V_{mean}=189$ km s$^{-1}$, $\sigma=67$ km s$^{-1}$; 
for all RRc stars,  $V_{mean}=186$ km s$^{-1}$, $\sigma=65$ km s$^{-1}$.

In Fig.~\ref{Fig17} we present radial velocity cumulative 
distributions of different LMC population as measured from our spectra.
The solid line represents LMC RR Lyrae stars, the thin line is for Mira stars.
Kolmogorov-Smirnov (KS) comparison of two data sets
gives the maximum difference between the cumulative distributions, 
$D = 0.26$ with a corresponding $P=0.20$. 
KS finds that the LMC RR Lyrae stars distribution is consistent
with a normal distribution
$P= 0.99$ where the normal distribution has mean $217.8\pm64$ km s$^{-1}$. 
KS finds that the
Mira stars distribution is consistent with a normal 
distribution $P= 0.64$ where the normal distribution has $229\pm51$ km s$^{-1}$.
The comparison between the 
$\omega$ Cen RR Lyrae stars and the 
LMC RR Lyrae stars shows 
the maximum difference between the cumulative distributions: $D$ is 0.5102 
with a corresponding $P$ of: 0.003. 
In the other words we have statistically significant difference 
with respect to 
$\omega$ Cen RR Lyrae stars and more data are necessary to confirm or reject
the hypotheses of difference between 
the LMC Mira and RR Lyrae stars.

 %
\begin{figure}[h]
\resizebox{\hsize}{!}{\includegraphics{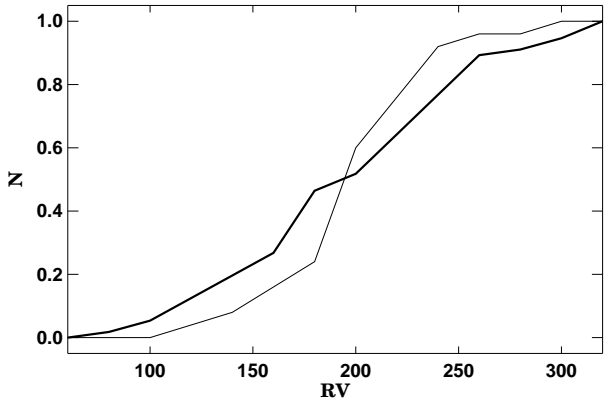}}
 \caption{The radial velocity
	cumulative distributions of different LMC population 
	as measured from our spectra.
	The solid line represents the LMC RR Lyrae stars, 
	the thin line is for Mira stars.}
\label{Fig17}
\end{figure}

In general, the amount of rotation compared with the velocity dispersion
determines whether a population is kinematically hot,
$V/\sigma<<1$, as opposed to kinematically cold, $V/\sigma\sim 1$.
The LMC RR Lyrae appear to belong to a kinematically hot population, 
the present data is insufficient to measure the rotation. The mean velocities
are less precise than the velocity dispersion, as seen by comparing
the different mean velocities of the Miras, RR Lyrae, and Cepheids. 
While there may be a hint of rotation for the RR Lyrae based on the sample
of the best 43 stars, this signature disappears when we include the extended
sample of 58 stars. 
In order to test how the rotation affects the velocity dispersion, we
computed the dispersions independently for the different fields. The results are 
given in Table~5. In column 2 we put the number ($N_{sel}$) of the  stars with the 
"best" radial velocity
measurements, columns 3 and 4 give their mean velocity ($RV$) 
and dispersion ($\sigma_{RV}$), while next
three columns $N_{all}$, $RV$ and $\sigma_{RV}$ are for the whole sample of stars.

\begin{table}[t]\tabcolsep=0.1pt\small
\begin{center}
\caption{Mean radial velocity and dispersion for several fields in LMC}
\label{Table5}
\begin{tabular}{c@{\hspace{0.5cm} }c@{\hspace{0.5cm} }c@{\hspace{0.5cm} }c@{\hspace{0.5cm} }c@{\hspace{0.5cm} }c@{\hspace{0.5cm}}c@{}}
\hline
\multicolumn{1}{c}{Field\hspace{0.5cm}}&
\multicolumn{1}{c}{$N_{sel}$\hspace{0.5cm}}&
\multicolumn{1}{c}{$RV$\hspace{0.5cm}}&
\multicolumn{1}{c}{$\sigma_{RV}$\hspace{1cm}}&
\multicolumn{1}{c}{$N_{all}$\hspace{0.5cm}}&
\multicolumn{1}{c}{$RV$\hspace{0.7cm}}&
\multicolumn{1}{c}{$\sigma_{RV}$}\\
\hline
10 &4 & 145& 45 & 5 & 179&85\\
11 &17& 241& 55 & 21&251&54\\
13 &5& 213& 20& 6&214&18\\
2 &4 &222&93&5&204&90\\
79& 7 &217 &42& 12&215&45\\
80& 7 &260 &25& 95 &274&37\\
\hline
\end{tabular}
\end{center}
\end{table}

The field with the largest number of RR Lyrae, Field 11, provides the
most accurate and stable measurement. The observed $\sigma=55$ confirms 
that the RR Lyrae have larger velocity dispersion than all other
measured populations, even though it is slightly smaller than the overall
dispersion observed, $\sigma=61$.
Clearly, more data are needed in order to estimate the rotation component. 

In addition, we must point out a difference in the published rotational
centers of the LMC, which bears on the measurement of the rotation of the
various populations. For example, van der Marel et al. (2002) obtain
$RA=5^h29\fm1$, $DEC=-69^\circ 47'$, while Alves \& Nelson (2000) obtain
$RA=5^h19\fm6$, $DEC=-69^\circ 58'$.
The RR Lyrae maps of Soszynski et al. (2003) indicate that the center of
the field RR Lyrae distribution in the LMC is located at:
$RA=5^h22\fm9$, $DEC=-69^\circ 39'$, offset from these previous determinations.

\section{Metallicities}

Preston's (1959) $\Delta S$ method is 
a classical method 
for determining RR Lyrae 
abundances from low resolution spectra. It measures the difference in spectral 
type between the CaII K line and the H ( $H_{\delta}, H_{\gamma}, H_{\beta}$) lines
relative to spectral type standards of approximately solar abundance.
Several modifications of the method can be found in the literature --
Butler (1975), Freeman \& Rodgers (1975), Smith (1995), Clementini at al. (2003).  
Layden (1994) adopted Freeman \& Rodgers (1975) method, which uses
the equivalent width (EW) of the CaII K-line (W(K)) and the  H-lines (W(H)) 
to estimate abundances but does not explicitly convert 
these 
measurements to spectral types. He derived the relation between W(K), W(H) and
metallicities [Fe/H] on the Zinn \& West (1984) scale for 142 field RRab Lyrae stars 
in our Galaxy. Equation 7 given in Layden (1994) is :
\begin{eqnarray*}
  W(K)&=& a+b\times W(H)+c \times [Fe/H]+ d\times W(H) \times [Fe/H] \hspace{10pt} \\
 \end{eqnarray*}
where W(K) and W(H)  are the  equivalent widths of CaII K-line and H-lines
respectively, and [Fe/H] is the metallicity in Zinn \& West (1984) scale.
The a, b, c, and d coefficients are given in Table~8 of the paper for three
different cases. The first one calibrates [Fe/H] vs. EW 
without correction for interstellar contribution to the K line.
To avoid the metal line contamination of $H_{\beta}$ in 
the second case, Layden (1994) uses only   
$H_{\delta}$ and $H_{\gamma}$ to measure
the EW of H-lines.
And finally in the third one he adds the $H_{\beta}$ line, after performing 
the corrections.   

Since we did not 
observe enough standard stars for calibration, 
we used the above
equation to determine the [Fe/H] of LMC RRab Lyrae stars.
Following Layden's (1994) three step approach, first we measured pseudo-equivalent widths of 
CaII K line and  $H_{\delta}, H_{\gamma}, H_{\beta}$ lines.
We used the SPLOT routine in IRAF, fitting the spectral 
lines with a Voigt profile.  
The pseudo-EWs were computed by dividing the instrumental fluxes within a small
spectral region centered on the selected feature, with the continuum. 
For RRab stars, the phases of rising light are to be avoided because of the effects of shock waves
and/or rapidly changing surface gravity. 
Using light curves from Kaluzny et al. (1997) for $\omega$Cen stars  and
from the MACHO database for LMC RR Lyrae's we calculated 
the phases of variable stars 
at the moment of our observations and 
rejected all the stars which had 
phases greater than 0.75. No correction for interstellar contribution to the K line was made, because the 
resolution of the spectra is not sufficient to separate stellar 
and interstellar component. To correct the metal line contamination of $H_{\beta}$ line, we
inspected the line profiles for signs of line doubling or for anomalously broad, shallow shapes.
In these cases we calculated [Fe/H], using only mean value of 
$H_{\delta}$ and $H_{\gamma}$ and  the second coefficients of Layden's calibration. 

We observed 17 RR stars in $\omega$Cen, but only 6 of them are RRab stars. The stars V88 and V111 
(Clement et al. 2001 classification) have no CaII K line in their spectrum, the star V144 was observed in two
masks, one of them is at phase 0.75. All these stars have [Fe/H] values 
measured by Butler et al. (1978) from low resolution spectra by classical $\Delta S$ method
and by Rey et al. (2000), which calibrated $hk$ index of the $Caby$ photometric system vs. metallicity.
The mean differences between  Rey et al. (2000) metallicities of these stars 
and our measurements is $0.11\pm0.32$, 
between Butler et al. (1978) measurements and ours is $0.01\pm0.07$,
Rey et al. (2000) and Butler et al. (1978)
metallicities have a differences  $0.10\pm0.36$. Taking into
account the small correction needed for the interstellar K line
we should expect sigma closer to 0.1. We have good agreement with spectroscopic measurements, but large
discrepancies with Rey et al.(2000) determination. They discuss 
the large difference in results from different methods and 
concluded that  $\Delta S$ metallicities of $\omega$Cen are not accurate. 
Of course we have only 4 stars,
but they support Jurcsik's (1998) suggestion, who also finds significant discrepancy between metallicities
obtained from Fourier decomposition of the light curves and the 
spectroscopic measurements.
Jurcsik's conclusion is that it is necessary to obtain high-accuracy spectroscopic observations in order 
to confirm and understand the nature of this difference.

The EW of the LMC RR Lyrae stars were 
measured independently by two of us (J.B. and M.R.)
and the unweighted mean value and the errors of the mean calculated. 
The average [Fe/H] value of our 23 RRab Lyrae stars is  [Fe/H] =$-1.50\pm0.1$ dex. 
The metallicity of another four stars are calculated from Fourier decomposition
of the light curves from Alcock et al. (2003) and are added to the spectroscopic measurements. 
As a result we have 27 metallicity determinations and the mean value is [Fe/H] =$-1.46\pm0.09$ dex.
The mean metallicity of RR Lyrae stars is in very good agreement with 
Clementini et al. (2003) average value [Fe/H] =$-1.48\pm0.29$ dex.
Thus, the field RR Lyrae stars seem to be more metal rich than 
globular cluster RR Lyrae star population (Walker 1992).
Our metallicities are on the Zinn \& West (1984) scale and 
they would be slightly different on the Carretta \& Gratton (1997) scale. 
This would introduce systematic
effects to the distance scale, that have been reviewed elsewhere (e.g. Walker 1998, Walker 2003).
Based on these 
considerations, we estimate a global uncertainty of 0.2 dex in the metallicity.

The individual [Fe/H] determinations are given in Table~6.
The distribution over the metallicity is shown on Fig.~\ref{Fig18}
and seems to fit well Gaussian distribution. Of course, more data
are necessary to confirm this.

%
\begin{figure}[h]
\resizebox{\hsize}{!}{\includegraphics{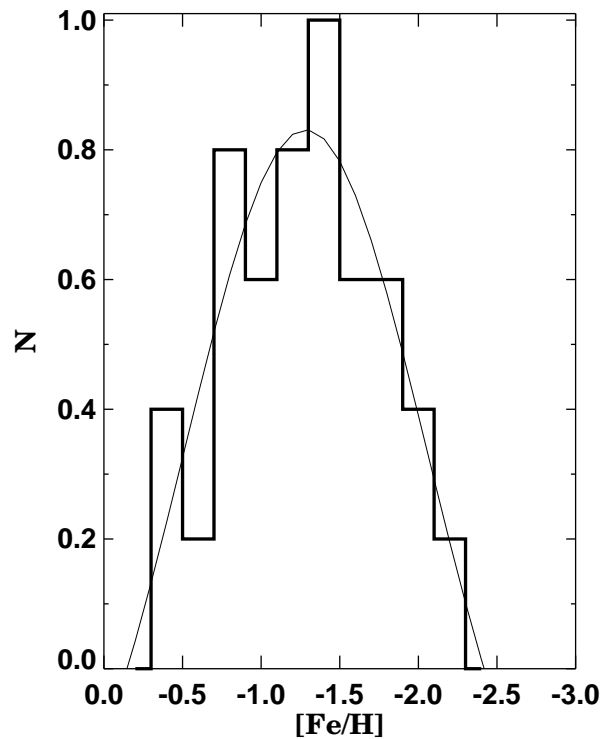}}
 \caption{ The metallicity histogram for RR Lyrae stars in our fields.
The solid line represents the Gaussian distribution.}
 \label{Fig18}
\end{figure}

\begin{table}[t]\tabcolsep=0.1pt\tiny
\begin{center}
\caption{Metallicities, $K$ magnitudes and $V-K$ colors of 
	the LMC RR Lyrae stars.}
\label{Table6}
\begin{tabular}{l@{ }c@{ }c@{ }l@{ }c@{ }c@{ }}
\hline
\multicolumn{1}{c}{MACHO name}&
\multicolumn{1}{c}{[Fe/H]}&
\multicolumn{1}{c}{$\sigma$}&
\multicolumn{1}{c}{Type}&
\multicolumn{1}{c}{$<K>$\hspace{0.5cm}}&
\multicolumn{1}{c}{$V-K$}\\
\hline
10.3802.311		&	$-$1.04&0.01	&	RRab	&	17.95	&	1.04	\\					
10.3802.339		&	$-$1.13&0.11	&	RRab	&	18.23	&	1.01	\\					
10.3802.446		&	$-$    		&		&	RRc	&	18.56	&	0.93	\\					
10.3922.978		&	$-$1.87&$-$		&	RRab	&	$-$	&	$-$	\\	 	 	 	 	 
10.3923.351		&	$-$    &		&	RRc	&	18.47	&	0.77	\\					
11.8622.757		&	$-$2.22&0.11	&	RRab	&	17.87	&	1.24	\\					
11.8623.3792		&	$-$1.46&0.05	&	RRab	&	$-$	&	$-$	\\					
11.8623.779		&	$-$1.65&0.01	&	RRab	&	17.96	&	1.24	\\					
11.8623.826		&	$-$0.73&0.02	&	RRab	&	18.01	&	0.94	\\					
11.8744.830		&	$-$2.14&0.01	&	RRab	&	$-$	&	$-$	\\					
11.8749.1208		&	$-$1.43&0.15	&	RRab	&	$-$	&	$-$	\\					
11.8749.1324		&	$-$2.33&0.06	&	RRab	&	18.13	&	1.49	\\					
11.8750.1425		&	$-$2.04&0.08	&	RRab	&	18.12	&	1.14	\\					
11.8750.1827		&	$-$2.10&0.18	&	RRab	&	18.42	&	1.33	\\					
11.8750.2045		&	$-$0.69&0.04	&	RRab	&	$-$	&	$-$	\\
11.8870.1275		&	$-$1.67&0.03	&	RRab	&	$-$	&	$-$	\\
11.8871.1362		&	$-$0.98&0.21	&	RRab	&	$-$	&	$-$	\\
13.5840.608		&	$-$1.08&0.06	&	RRab	&	$-$	&	$-$	\\
13.5840.730		&	$-$1.20&0.19	&	RRab	&	$-$	&	$-$	\\
13.5840.768		&	$-$	 &		&	RRab	&	18.20	&	1.45	\\
13.5961.435		&	$-$	 &		&	RRab	&	18.24	&	0.60	\\
13.5961.511		&	$-$	 &		&	RRab	&	17.66	&	1.44	\\
13.5961.623		&	$-$	 &		&	RRab	&	18.06	&	1.18	\\
13.5961.648		&	$-$	 &		&	RRab	&	18.08	&	1.31	\\
13.5961.720		&	$-$	 &		&	RRab	&	18.13	&	1.29	\\
13.5962.547		&	$-$1.31&0.04	&	RRab	&	$-$	&	$-$	\\
13.5962.656		&	$-$    &		&	RRab	&	17.90	&	1.44	\\
2.5507.6046		&	$-$	 &	 	&	RRab	&	18.07	&	1.55	\\		
2.5507.6257		&	$-$	 & 		&	RRc	&	18.22	&	1.53	\\		
2.5627.4847		&	$-$	 &		&	RRc	&	19.22	&	0.76	\\		
2.5628.5690		&	$-$	 &		&	RRab	&	18.89	&	0.80	\\		
2.5628.6276		&	$-$0.57&0.01	&	RRab	&	18.43	&	1.39	\\		
79.5507.1485		&	$-$1.85&0.11	&	RRab	&	$-$	&	$-$	\\		
79.5507.1580		&	$-$    &		&	RRab	&	18.03	&	1.43	\\		
79.5508.427		&	$-$1.66&0.17	&	RRab	&	$-$	&	$-$	\\		 
79.5508.534		&	$-$1.31&0.04	&	RRab	&	$-$	&	$-$	\\		 
79.5508.682		&	$-$1.92&0.06	&	RRab	&	$-$	&	$-$	\\		
79.5508.735		&	$-$1.51&0.13	&	RRab	&	$-$	&	$-$	\\		
79.5627.1761		&	$-$	 &		&	RRc	&	19.11	&	0.62	\\		
79.5628.1065		&	$-$1.24&$-$	&	RRab	&	$-$	&	$-$	\\		
79.5628.1300		&	$-$    &		&	RRab	&	18.02	&	1.40	\\		
79.5628.1650		&	$-$	 &		&	RRc	&	17.98	&	1.24	\\
79.5628.2110		&	$-$	 &		&	RRab	&	18.28	&	1.52	\\
80.6468.2582		&	$-$	 &		&	RRab	&	18.17	&	1.44	\\
80.6468.2616		&	$-$	 &		&	RRc	&	17.99	&	1.22	\\
80.6468.2765		&	$-$	 &		&	RRc	&	18.26	&	0.97	\\
80.6468.2799		&	$-$	 &		&	RRab	&	17.76	&	1.64	\\
80.6468.3667		&	$-$	 &		&	RRc	&	18.51	&	1.54	\\
80.6469.1657		&	$-$1.90&0.09	&	RRab	&	18.00	&	1.25	\\
80.6469.1712		&	$-$-1.61&0.01	&	RRab	&	$-$	&	$-$	\\
80.6588.2703		&	$-$0.91&0.16	&	RRab	&	$-$	&	$-$	\\
\hline
\end{tabular}
\end{center}
\end{table}

It is well known that the absolute magnitude of an RR Lyrae star, $M_{V}$, is 
a function of metallicity [Fe/H]
(Sandage 1981). Recently, Olech et al. (2003), using new photometry 
of RRab Lyrae stars in 
$\omega$Cen, derived the relation in the form : 
$M_{V}=0.26\times \mathrm{[Fe/H]}+0.89 $. 
Despite the large scatter, in general our data agree with this relation as can be seen in 
Fig.~\ref{Fig19}. To calculate $M_{V}$, we used 
the LMC distance modulus of 18.48 determined in Sec.~ 3 and 
the reddening of E(B-V)=0.11 (Clementini et al. 2003).

%
\begin{figure}[h]
\resizebox{\hsize}{!}{\includegraphics{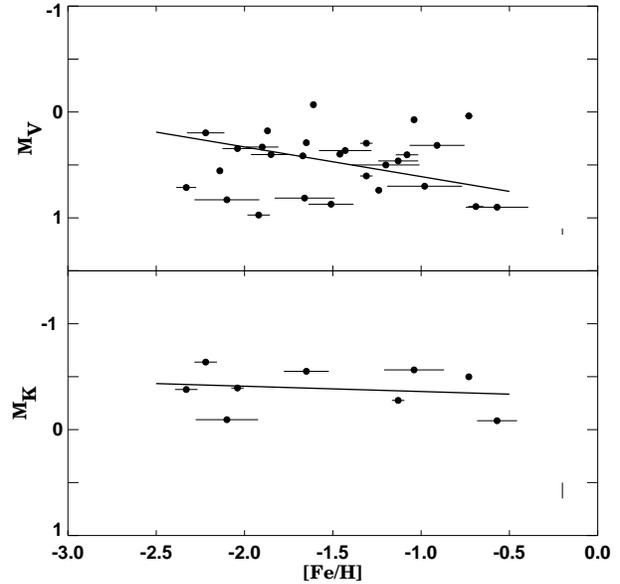}}
 \caption{Top panel: $M_{V}$,[Fe/H] plot for RRab stars in LMC. The
Olech et al (2003) relation is shown with the solid line. Bottom panel:
 $M_{K}$,[Fe/H] plot for RRab stars in LMC. 
	The solid line is the best linear fit. The vertical lines
	are the global error of the  $M_{V}$ and $M_{K}$ magnitudes. 
 The adopted distance modules and reddening are 18.48 and 0.11,
respectively (see text).}
\label{Fig19}
\end{figure}

In the $M_{K}$,[Fe/H] plot, however, the relation is flatter than in the
V band as can be seen on the bottom panel of Fig.~\ref{Fig19}. 
The best fit of our 9 points gives:  $M_{K}=0.05\times[Fe/H]-0.31$ dex. 
There is an indication that the absolute K magnitude is 
less sensitive to the metallicity.

\section{Conclusions}
We measure a velocity dispersion of $\sigma=53\pm 10$ km s$^{-1}$ for 43 LMC RR Lyrae.
This dispersion indicates a kinematically hot population, similar
to the Milky Way, where the RR Lyrae are distributed in the halo.
This velocity dispersion is also larger than that of the old LMC globular
clusters.

We measure a mean $[Fe/H]= -1.46\pm 0.09$ dex based on $N=27$ field LMC RR Lyrae.
This measurement agrees with the recent result of Clementini et al. (2003),
who found $[Fe/H]= -1.48\pm 0.29$ dex based on $N=100$ field RR Lyrae in the LMC.
This mean metallicity is different from the globular clusters: Walker (1992) 
finds $[Fe/H]= -1.9\pm 0.2$ dex based on $N=182$ RR~Lyrae in seven LMC globular clusters.
This difference, coupled with the difference in the kinematics and the period
distributions, suggests that
field RR Lyrae and globular clusters trace different populations in the LMC.
 
We measure the mean $K$-band magnitudes of 37 field RR Lyrae in the LMC, 
$K=18.20\pm 0.06$. Based on this mean magnitude and on the theoretical near-IR
period-luminosity-metallicity relations of Bono et al (2001, 2003), we compute 
an LMC distance modulus $\mu_0 = 18.48\pm 0.08$.

\begin{acknowledgements}
JB, DM, and MR are supported by Fondap Center for Astrophysics 15010003.
KHC's work was performed under the auspices of the U.S. Department of Energy,
National Nuclear Security Administration by the University of California, 
Lawrence Livermore National Laboratory under contract No. W-7405-Eng-48.
The authors gratefully acknowledge the comments by the referee Dr. Bono
and comments by M.~Catelan and H.~Smith.
\end{acknowledgements}

\end{document}